\documentclass[fleqn,10pt]{wlscirep}
\usepackage[T1]{fontenc}
\usepackage{graphicx}
\usepackage{dcolumn}
\usepackage{bm}
\usepackage{mathrsfs}
\usepackage[mathlines]{lineno}
\usepackage[utf8]{inputenc}
\usepackage{soul,color,xcolor}
\usepackage{textcomp}

\title{Evolution of quantum criticality in underdoped cuprates}

\author[1]{Changshuai Lan}
\author[1,3,*]{Chengyu Yan}
\author[1]{Yihang Li}
\author[1]{Qiao Chen}
\author[1]{Huai Guan}
\author[1]{Xinming Zhao}
\author[2]{Dong Wu}
\author[1]{Butian Zhang}
\author[1]{Youwei Zhang}
\author[1,3,*]{Shun Wang}

\affil[1]{Ministry of Education Key Laboratory of Fundamental Physical Quantities Measurement and Hubei Key Laboratory of Gravitation and Quantum Physics, National Precise Gravity Measurement Facility and School of Physics, Huazhong University of Science and Technology, Wuhan 430074, P. R. China}
\affil[2]{Beijing Academy of Quantum Information Sciences, Beijing 100193, P. R. China}
\affil[3]{Institute for Quantum Science and Engineering, Huazhong University of Science and Technology, Wuhan 430074, P. R. China}

\affil[*]{corresponding author(s): Chengyu Yan(chengyu$\_$yan@hust.edu.cn), Shun Wang(shun@hust.edu.cn)}

\begin{abstract}
Quantum criticality, with both static and dynamic information of the system intrinsically encoded in characteristic length scales, serves as one of the most sensitive and universal probes to monitor quantum phase transition. Qualitatively different quantum criticality behaviours have been widely observed even in the same condensed matter system. The discrepancy is attributed to sample specificity but has not been systemically addressed. Here we report a single-parameter driven three-stage evolution of quantum criticality unveiled in superconductor-insulator transition in underdoped ${\mathrm{Bi}}_{2}{\mathrm{Sr}}_{2}{\mathrm{CaCu}}_{2}{\mathrm{O}}_{8+\ensuremath{\delta}}$ flakes. The evolution starts with a single quantum critical point emerging at the boundary between the superconducting and antiferromagnetic phases, then evolving into anomalous quantum Griffiths singularity at the medium doping levels and eventually being replaced by quantum Griffiths singularity in the deep superconducting regime. A puddle model that incorporates the developments of antiferromagnetic correlation can capture the evolution. The results offer a new aspect to examine previous seemingly sample-specific quantum critical behavior and lay the foundation for further exploring complex quantum criticality in strongly correlated systems; meanwhile they shed light on the detailed interaction between superconductivity and antiferromagnetism in cuprates.

\end{abstract}

\begin{document}                              
\flushbottom
\maketitle

\thispagestyle{empty}


 
Quantum criticality, associated with the divergence of spatial correlation length $\xi_{l}\propto\left|\delta\right|^{-\nu}$ and temporal correlation length $\xi_\tau\propto\left|\delta\right|^{-\psi}$, serves as a universal probe to characterize quantum phase transition in various condensed matter systems, such as unconventional superconductivity\cite{IntroFe,monolayerBSCCO,QCPinCu2,QCPinWTe}, heavy-fermions\cite{IntroHF1,IntroHF2},complex magnetism\cite{IntroFM1}, and topological states\cite{IntroinTP,IntroinTP2}. Recently, a series of research on quantum criticality even showcase the outstanding insights about the consistency of physical mechanisms from nuclei\cite{Intronuclei} to the cosmos\cite{Introdarkmatter,Strangerthanmetal}.

Depending on the details of how the system approaches the quantum critical point and the stability of the quantum critical point determined by the Harris criterion, the quantum criticality behavior, after taking quenched disorder into account, can exhibit rather non-trival\cite{Harris_1974,harris}. If Harris criterion is fulfilled $dv>2$, where $v$ is the correlation length exponent and $d$ is the dimensionality, a single quantum critical point (QCP) is expected. QCP is a fixed critical point against external control parameters in a quantum phase transition process\cite{QCPinInO,QCPinBi,QCPinLSCO,monolayerBSCCO}. If Harris criterion is violated $dv<2$, quantum Griffiths singularity (QGS) will occur. The primary experimental features of QGS are the varying critical points and divergence of the critical exponent\cite{QGSinGa,QGSinionicgating,QGSinPdTe,QGSinNiO}. Further, if QGS is supplemented with the competition between different interactions or correlated states, such as that between spin-orbit coupling and superconducting pairing, the anomalous quantum Griffiths singularity (AQGS) may emerge. The signatures of AQGS are divergence of the critical exponent along with a nonmonotonic phase boundary\cite{AQGS,AQGS2}. Currently, it is believed that the core mechanisms of QGS and AQGS can be converted into each other\cite{AQGS2}, but one with monotonic phase boundary and the other with nonmonotonic one. The understanding of AQGS is still in its infancy. Nevertheless, the observation of behavior that deviates from standard QCP alone has constituted a leap in the understanding of quantum criticality. 

However, there are fundamental issues yet to be addressed. Since disorder and quantum fluctuations have profound influences on quantum criticality, it has always been noticed that large variation in the critical exponents in the same condensed system prepared with different methods or even the same methods in experiments\cite{monolayerBSCCO,IonicBSCCO,zvinBi,zvinPCCO}. More importantly, the qualitative behavior of quantum criticality can also be different from sample to sample for the same system, for instance, QCP and QGS behavior have been independently observed in indium oxide films\cite{QCPinInO,goldman} and tin nanoisland-arrays on graphene\cite{DQCPingr,QGSingr}. Somehow, coexistence of multiple quantum criticality behavior in a same sample  has not been firmly established and leaves alone commuting different quantum criticality behaviors via continuous modulation in parameter space. It is not sure that the aforementioned difference in quantum criticality is sample-specific, or that they can be clips of a more ergodic evolution that involves multiple quantum critical behaviors. To resolve this fundamental puzzle in quantum criticality, it is highly desirable to demonstrate the coexistence of several types of quantum criticality in a single system and drive a continuous transition between these quantum critical behaviors via a single control knob. Meanwhile, the evolution of quantum criticality naturally monitors the impact of the control parameter in synchronizing disorder dominated local dynamics, captured by the local temporal correlation length associated with the disorder, in the case of QGS or AQGS\cite{reviewofQGS} to a global dynamics, embedded in the global temporal correlation length for the entire system, in the case of QCP\cite{ReviewQPT2}. This synchronization process is of general interest across different branches of physics and possibly unveils novel dynamics of phases beyond the standard framework of phase transitions\cite{reviewofQGS,ReviewQPT2}. Also, if it is indeed possible to drive an evolution of quantum criticality, then quantum criticality can be implemented as a universal probe to scan the ground states of a system in a wide range of parameter space. 
         
In this work, we use high temperature superconductor ${\mathrm{Bi}}_{2}{\mathrm{Sr}}_{2}{\mathrm{CaCu}}_{2}{\mathrm{O}}_{8+\ensuremath{\delta}}$ (BSCCO) as a model system to demonstrate the transition between different quantum criticality driven by doping level. Combining the enhanced quantum fluctuations in two-dimensional cuprates, the complicated competition between different correlated states, especially the one between superconductivity and antiferromagnetism, and the ability to modulate the intensity and distribution of disorders in a wide range, we have unveiled an evolution of quantum criticality with increasing doping levels in underdoped BSCCO. QCP emerges at the boundary between the superconducting and antiferromagnetic phases, then evolves into AQGS at the medium doping levels, and is eventually replaced by QGS in the deep superconducting regime within the superconducting dome. A puddle model that incorporates the developments of antiferromagnetic correlation is proposed to capture the observations. Therefore, our work highlights the possibility of multiple-stage evolution quantum criticality driven by a single control parameter, and thus starts a new venture for exploring complex quantum criticality in strongly correlated systems. Besides, the AQGS and QCP observed in the work not only provide a sensitive probe to determine the doping range for the emergence of antiferromagnetic correlation in the superconducting dome, but also shed light on the detailed interaction between superconductivity and antiferromagnetism. The above two points are vital in identifying the mechanism of the occurrence of superconductivity from the parent antiferromagnetic insulator in cuprates.

\section*{Modulation and characterization of samples}

Intensity and distribution of disorder is one of the major knobs to determine different quantum criticality behavior\cite{Harris_1974,Lin1,Lin2}. It is a well-established technique that induces disorder by tuning the doping level via vacuum annealing\cite{monolayerBSCCO,ourwork}. In this work, the BSCCO flakes were vacuum annealed at or above 300K. We can use reentrant superconductivity behavior arising from disorders in ab-plane to quantify the doping level\cite{reentrant,ourwork}. From the nonmonotonic RT curves depicted in Fig. \ref{1}(a), we can conclude two critical temperatures. One is the trivial critical temperature $T_{c,max} \approx 88K$, which is independent with doping level, and another is the second critical temperature $T_{c}$, where the resistance drops to zero again and thus signifies reentrant superconductivity. The doping level $p$ can be extracted from $T_{c}$ by the empirical relation: $T_{c}=T_{c,max}\left[{\mathrm{1}}-{\mathrm{82.6}}\left(p-{\mathrm{0.16}}\right)^{2}\right]$\cite{SITofBSCCO,monolayerBSCCO}. It is found that the doping level spans from 0.12 to as low as the complete insulating as summarized in Fig.\ref{1}(b).

Superconductor-insulator transition (SIT), as a paradigm of quantum criticality, can be observed in the entire doping range. Fig.\ref{1}(c)-(e) shows typical behavior of temperature-dependent resistance under out-of-plane magnetic fields at three different doping levels, $p=0.073$ (QGS), $p=0.055$ (AQGS) and $p=0.0525$ (QCP), respectively. The temperature-dependent resistance curves are qualitatively similar in all three cases. First of all, $T_{c}$ and the resistance peak for reentrant superconductivity shift towards lower temperatures with increasing magnetic field, as results of vortex dynamics\cite{ourwork}. More importantly, the resistance rises up with decreasing temperature once the magnetic field exceeds a doping-level specific value, suggesting the occurrence of SIT\cite{QCPinBi,QCPinInO,Lin1,Lin2}. Hereafter, we will focus on magneto-resistance(referred to as RH) at relatively low temperature (lower than the doping dependent $T_{c}$) to highlight different quantum criticality behaviors emerging in SIT and unveil the transition between these behaviors.

\section*{Evolution of quantum criticality}

We have observed an evolution of quantum criticality driven by doping level. It starts with QCP, then transfers into AQGS and seemingly ends as QGS when the doping level gradually increases from the boundary between the superconducting and antiferromagnetic phases to the deep superconducting phase in the underdoped side of the superconducting dome. It should be emphasized that magnetic field drives SIT where quantum criticality behavior emerges, while doping level determines the type of criticality behavior, which is different from previous studies\cite{monolayerBSCCO,IonicBSCCO}.
 
At the boundary between the superconducting and antiferromagnetic phase, $p = 0.0525$, QCP emerges. The RH data for the out-of-plane magnetic fields at $p = 0.0525$ cross at a single critical point ($B_{c}=100$ mT) in the entire temperature range as shown in Fig.\ref{2}(a). The temperature range is chosen according to the SIT plateau in the corresponding temperature-dependent resistance. We conduct finite-size scaling (FSS) analysis on the RH data to further validate the quantum criticality behavior: RH follows the scaling law $R=R_{c}\cdot F(|B-B_{c}|T^{-1/z\nu})$ in the vicinity of $B_{c}$. Here, $R_{c}$ and $B_{c}$ are the resistance and the magnetic field of the critical point, $F$ is an arbitrary function with $F(0)=1$, and $z\nu$ is the critical exponent. By scaling the magnetic field as $|B-B_{c}|t$, where $t=(T/T_{0})^{-1/z\nu}$ with $T_{0}$ the lowest temperature, normalized RH ($R/R_c$) should collapse together at a quantum critical point. The gradient of $T/T_{0}$ versus $t$ in logarithmic coordinates is the critical exponent $z\nu$ (details can be seen in the Section 2 of the Supplement Information). It is clear that RH data in the entire temperature range collapse satisfactorily after scaling, as shown in Fig.\ref{2}(b). Interestingly, the scaling parameter $t$ shows a piece-wise linear behavior against $T/T_{0}$ as highlighted in Fig.\ref{2}(c).  $z\nu$ is about 1.5 below 4K and 0.4 above 4K. Such a transition is a result of competition between phase fluctuation and disorder\cite{DQCPinOxide,DQCPingr,monolayerBSCCO}, which will be elaborated on later on. 

AQGS has been monitored within $0.055 < p < 0.06$. Fig.\ref{3} shows the typical result at $p = 0.055$ (data for AQGS at other doping levels can be found in the Section 4 of the Supplement Information). Here, RH curves for two adjacent temperature crosses at a given magnetic field, as shown in Fig.\ref{3}(a) and its inset. To determine if there is quantum criticality behavior, we adapt the standard protocol of analyzing data with varying critical points\cite{QGSinGa,QGSinionicgating,QGSinPdTe,AQGS,QGSingr}. We divide the data into a series of small temperature windows and then conduct FSS analysis in each window. In each window, a cross point $B_{c}$ can be identified and the RH data collapses in the vicinity of this $B_{c}$ after scaling. Fig.\ref{3}(b) indicates that $B_{c}$ initially increases by lowering the temperature and then gradually sinks at the lowest temperature regime. The critical exponent $z\nu$ follows $z{\nu}\propto(B-B_{c}^{*})^{-0.6}$ in the $B_{c}$-decreasing segment as shown in Fig.\ref{3}(c). The nonmonotonic $B_{c}-T$ phase boundary and scaling exponents of -0.6 are in good agreement with features of AQGS reported in recent research. It is suggested that suppression of superconductivity due to strong spin-orbit coupling together with quenched disorder leads to AQGS\cite{AQGS,AQGS2}. However, underdoped BSCCO is not known for strong spin-orbit coupling capable of suppressing superconductivity\cite{SOCincuprates}. Instead, we believe the competition between superconductivity and antiferromagnetism in the low doping regime, which naturally suppresses superconductivity at low temperature\cite{afmwithsc}, is likely to result in AQGS in the presence of quenched disorder. This assertion would be justified in the discussion section. 

Increasing the doping level even further ($0.06 < p < 0.09$), it seems that QGS takes over.  Fig.\ref{4} shows the typical result at $p = 0.073$ (data for QGS at other doping levels can be found in the Section 4 of the Supplement Information). The RH data is similar to that of AQGS in terms of varying crossing points. More insights can be unearthed by extracting cross point $B_{c}$ with FSS. $B_{c}$ soars with decreasing temperature up to the lowest available temperature in our study and closely resembles the trend of vortex melting field in BSCCO\cite{ourwork}, as highlighted in Fig.\ref{4}(b). It is clear that $z{\nu}$ diverges at zero temperature and follows the activating-type scaling $z{\nu}\propto(B_{c}^{*}-B)^{-\nu\psi}$. $\nu\psi$=0.6, signifying 2D infinite-randomness, has been universally observed for QGS in a wide range of 2D superconducting systems\cite{reviewofQGS}. Hence, we have identified both primary features of QGS, the varying quantum critical points and divergence of $z\nu$, in this regime. It is necessary to stress that due to the limitation of the base temperature of the cryostats, we can not directly exclude the possibility that nonmonotonic $B_{c}$, the signature for AQGS, may occur at extremely low temperature. In other words, the exact doping level boundary between QGS and AQGS may need more advanced measurements. However, QGS should take over AQGS at high doping levels due to the vanishing antiferromagnetic fluctuation in the deep superconducting phase.

\section*{Discussion}

The development of antiferromagnetic correlation across superconducting puddles can qualitatively explain the evolution of quantum criticality against the doping level.

The puddle model is a paradigm to understand the impact of local microstructures on global properties in cuprates\cite{puddle2,monolayerBSCCO,monolayerBSCCO2,stmofcuprates}. It suggests that vacancies and concentrations of oxygen will fragment the system into superconducting puddles with different $T_c$. The superconducting puddles are surrounded by non-superconducting regions of various sizes. The size and spatial distribution of non-superconducting regions can regulate global superconductivity via tuning the Josephson coupling between superconducting puddles\cite{puddle1}. In addition, the nature of the non-superconducting region, i.e., being a trivial normal state core or some sorts of local short-range fluctuations or short-range orders, gives more flavor to the puddle model to elucidate a wide range of phenomena in cuprates. Then we can now apply the general idea of the puddle model along with the characteristics of non-superconducting regions inferred from literature to different doping ranges.        

\begin{figure}[bt]
	\centering
	\includegraphics[width=1\textwidth]{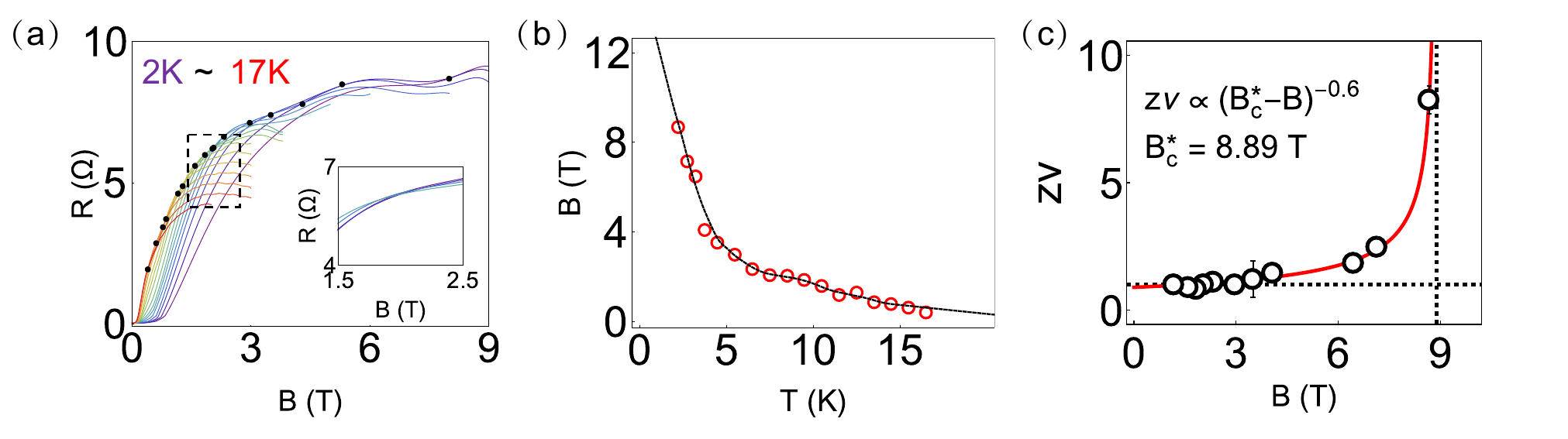}
	\caption{QGS in BSCCO at p = 0.073. (a) Magneto-resistance with temperature increasing from 2 K to 17 K by 1 K steps. The black circles mark the cross points of every two curves at adjacent temperature. The inset shows a zoom-in of data in the vicinity of a typical cross point ( from 8 K to 10 K by 0.5 steps), which is marked on the main plot with a dashed box. Details on finite-size scaling analysis for QGS can be found in the Section 2 of the Supplement Information. (b) Temperature dependence of cross points $B_{c}$. The black curve guides the trend of $B_{c}$. (c) Dynamical exponent $z{\nu}$ as a function of magnetic field. The red curve shows the fitting based on the activated scaling law. The horizontal dashed line shows $z{\nu} = 1$ and the vertical one marks the extracted $B_{c}^{*}$.}
	\label{4}
\end{figure}

\begin{figure}[h]
	\centering
	\includegraphics[width=0.95\textwidth]{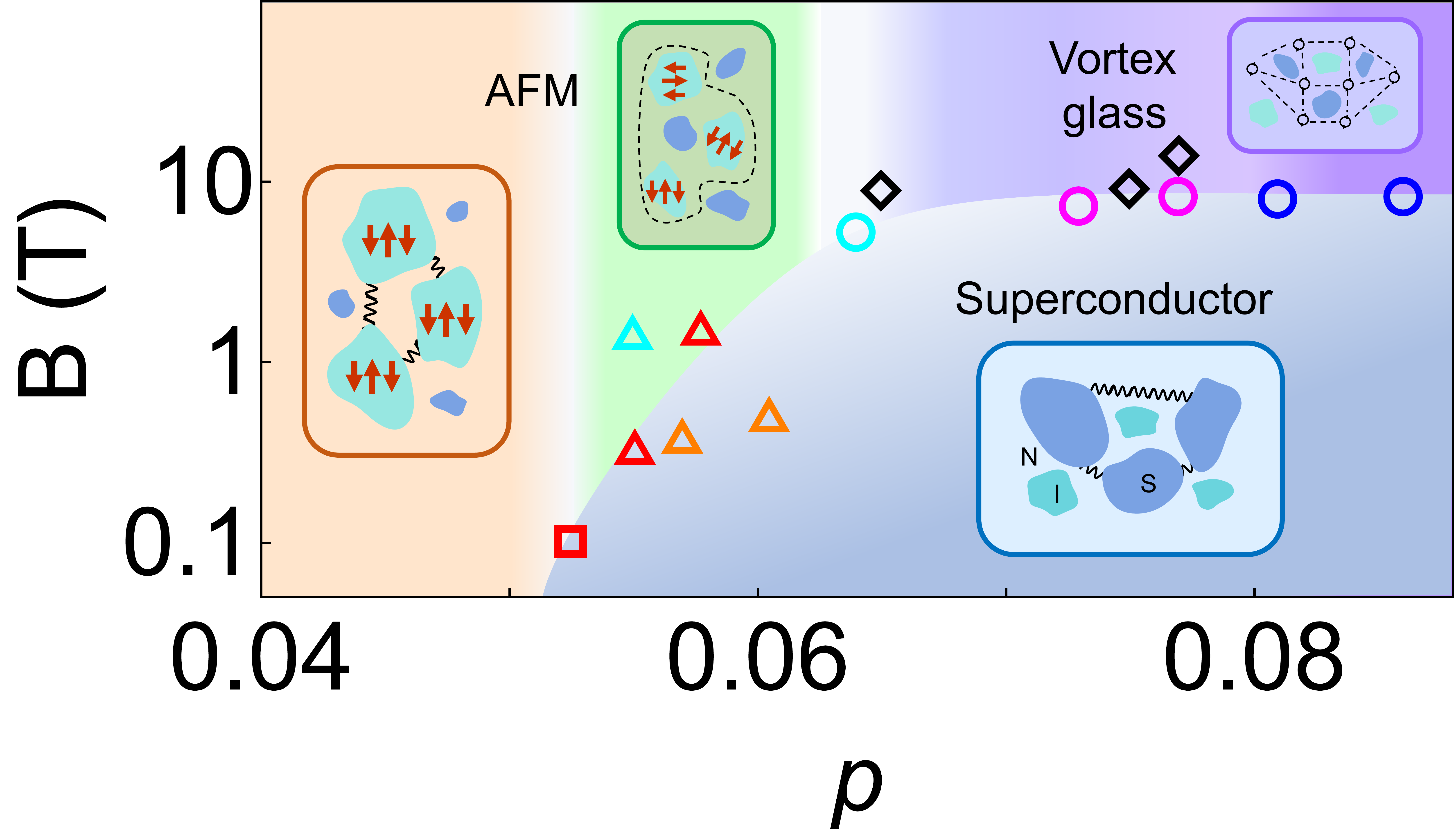}
	\caption{Field-doping phase diagram for the evolution of quantum criticality. The blue region is superconductor, the purple region is vortex glass, the green region is insulator with local antiferromagnetic orders and the orange region is insulator with long-range antiferromagnetism phase coherence. The phase boundary is decided by $B_{c}^{*}$ of different quantum criticality. Circle markers show $B_{c}^{*}$ of QGS. Triangle markers show $B_{c}^{*}$ of AQGS. $B_{c}^{*}$ of QGS and AQGS are extracted from FSS scaling analysis. Square markers show $B_{c}$ of QCP. Points with the same color code are acquired from the same sample. Black diamond markers represent the vortex melting field at 0 K from previous works\cite{ourwork}. The insets show the illustration of phases contributing to quantum criticality. The blue puddles with S denote superconductor. The cyan puddles with I depict insulator. The black circles with arrows show vortices. The red arrows mean the antiferromagnetic orders. Quenched disorders decompose system into Josephson coupled superconducting puddles that exhibit superconductivity global-wise at zero magnetic field. Increasing out-of-plane magnetic field will drive a quantum phase transition from this global superconducting phase into various phases. At high doping levels, QGS occurs between the global superconducting phase and the vortex glass phase. At medium doping levels, AQGS occurs between the global superconducting phase and the uncoupled superconducting puddles separated by normal regions with local antiferromagnetic fluctuation. The local antiferromagnetic fluctuation is responsible for non-monotonic field-temperature phase boundary. At low doping levels, QCP occurs between the global superconducting phase and the global antiferromagnetic phase.
}
	\label{5}
\end{figure}

QGS regime ($0.06 < p < 0.09$): In the deep superconducting regime of the underdoped side. The non-superconducting regimes are primarily constituted by trivial normal state cores, acting as quenched disorder in quantum phase transition\cite{harris,Harris_1974}, and decompose the system into superconducting puddles. The superconducting puddles are coupled together via Josephson coupling\cite{puddle1} and behave phase coherence globally at low temperature, as shown in the blue region of Fig.\ref{5}. Increasing the magnetic field will weaken the Josephson coupling and also induce a vortex glass state (the purple region in Fig.\ref{5}). Hence, global superconductivity ceases to exist, whereas the ordered superconducting phase is restricted within the uncoupled superconducting puddles. Each superconducting puddle dictates the quantum phase transition in the vicinity of its critical point, resulting in the varying critical points. Therefore, QGS emerges in this regime. Since vortex glass plays a central role in SIT in this regime, it is no wonder that the phase boundary, $B_c$-T curve in Fig.\ref{4}(b), emulates the vortex-melting field curve. The same picture has been successfully implemented in a wide range of materials to explain the occurrence of QGS\cite{QGSinGa,QGSinionicgating,QGSinPdTe,QGSinNiO}.

\begin{figure}[h]
	\centering
	\includegraphics[width=1\textwidth]{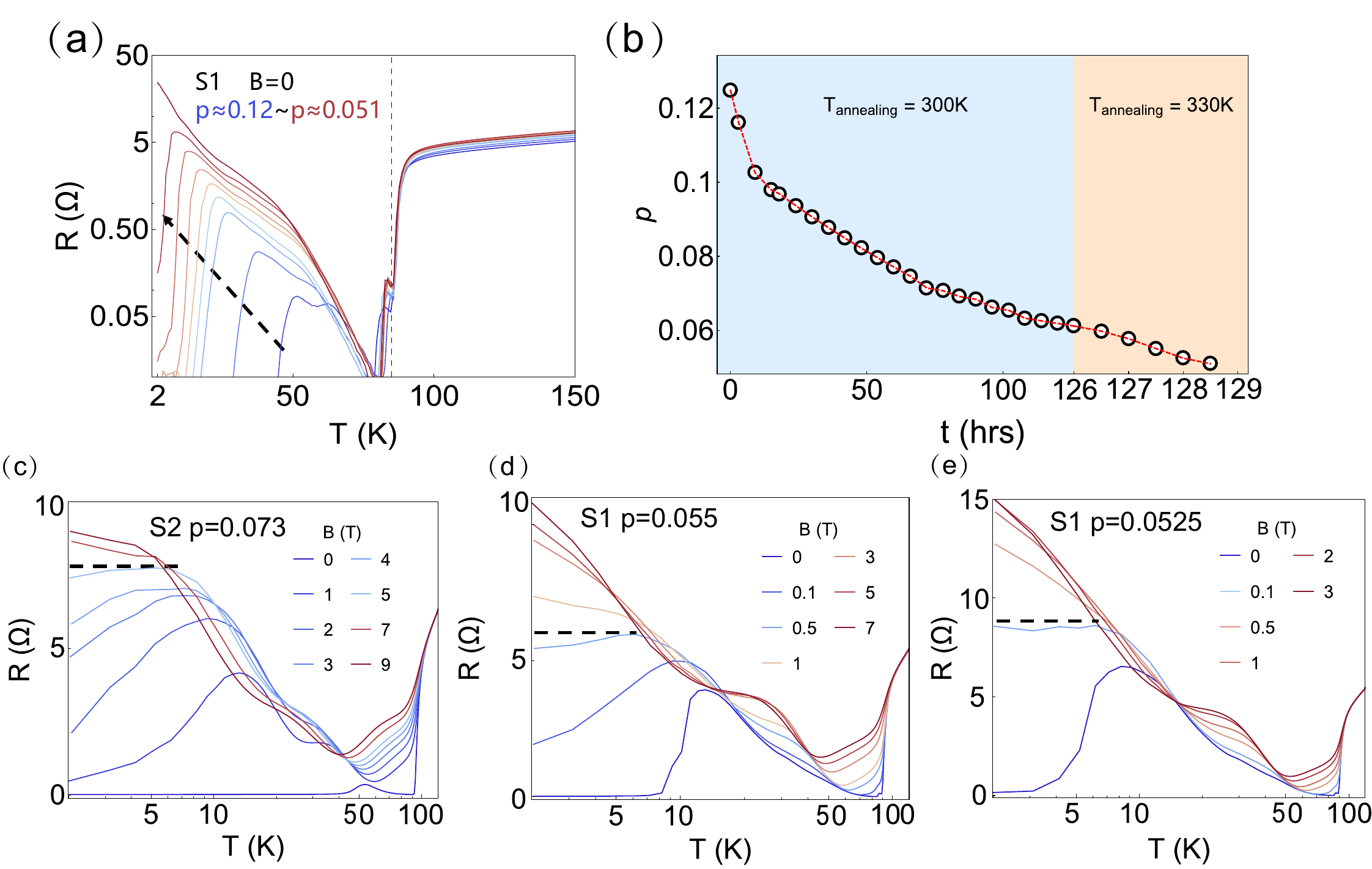}
	\caption{Characterization of vacuum annealed devices. (a) Temperature dependent resistance at different doping levels. The black dashed line guides the doping-level insensitive critical temperature $T_{c,max}$. The black arrow guides the critical temperature $T_{c}$ that depends on the doping level. (b) Doping level as a function of annealing time. Oxygen releasing rate is significantly reduced when doping level approaches $p=0.06$ at a annealing temperature of 300 K, therefore the annealing temperature is incremented to 330 K to boost the releasing process and hence obtain lower doping level. (c)-(e) Temperature dependent resistance in the presence of out-of-plane fields for $p=0.073$, $p=0.055$ and $p=0.0525$, respectively. The black dashed lines hint the onset of superconductor-insulator transition.}
	\label{1}
\end{figure}

AQGS regime ($0.055 < p < 0.06$): Upon lowering the doping level, the main picture remains, which is responsible for the divergence of the dynamical critical exponent. However, it requires an additional knob to generate the nonmonotonic $B_c$-T phase boundary that differs AQGS from QGS. In previous studies carried out in disordered metal film and interface superconductor, spin-orbit coupling can result in the nonmonotonic phase boundary by suppressing superconductivity\cite{AQGS,AQGS2}. Underdoped BSCCO, on the other hand, is not equipped with sizeable spin-orbit coupling\cite{SOCincuprates}. Instead, numerous ARPES\cite{afmincuprates2}, magnetoresistance\cite{afmincuprates} and neutron-scattering\cite{localafm3} measurements suggest the emergence of local antiferromagnetic fluctuation or local AFM order within the non-superconducting region at low doping level in cuprates. The local AFM component can potentially suppress superconductivity similar to what happens in antiferromagnetic superconductors such as ${\mathrm{Ho}}{\mathrm{Ni}}_{2}{\mathrm{B}}_{2}{\mathrm{C}}$\cite{afmwithsc}. More interestingly, the nonmonotonic $B_c$-T phase boundary in antiferromagnetic superconductors actually manifests itself as reentrant superconducting behavior\cite{afmwithsc}, which also occurs in our experiment as shown in Fig.\ref{1}. Inserting the characteristics of non-superconducting into the phase transition process, it is seen that increasing magnetic field will drive a quantum phase transition from an ordered phase with global superconductivity into a disordered phase constituted by rare regions with local antiferromagnetism as depicted in the green region of Fig.\ref{5}. The puddle-like distribution of the non-superconducting region results in the divergence of the dynamical critical exponent, meanwhile the local-scale AFM characteristic of the non-superconducting region leads to the nonmonotonic $B_c$-T phase boundary. Therefore, AQGS can be observed in this regime. Antiferromagnetic fluctuation naturally diminishes by increasing the doping level into the deep superconducting regime, therefore QGS can take place over AQGS in that regime.

\begin{figure}[hb]
	\centering
	\includegraphics[width=1\textwidth]{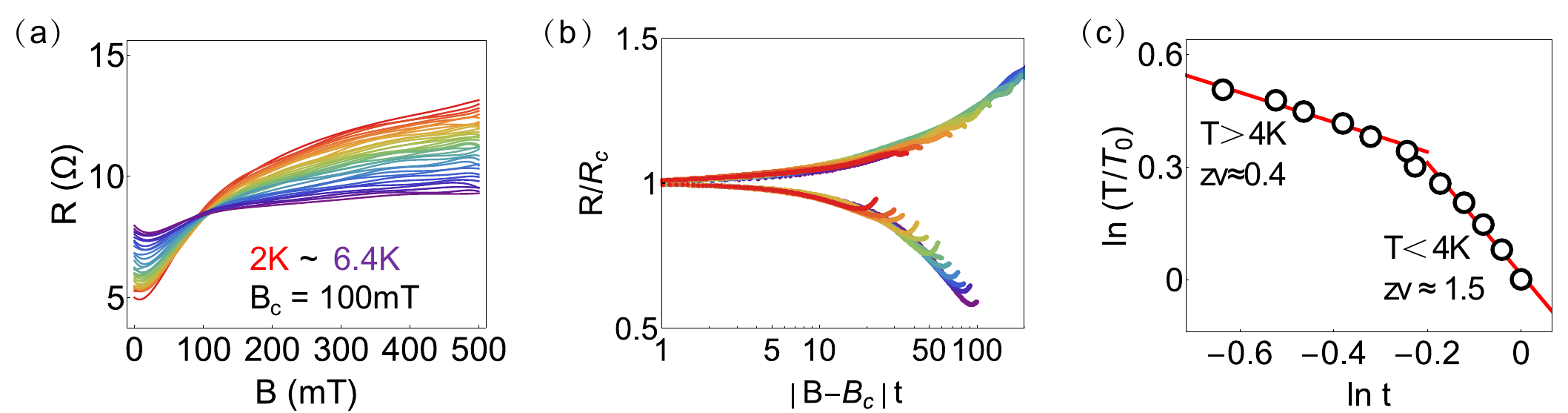}
	\caption{Signatures of QCP in BSCCO at p=0.0525. (a) Magneto-resistance with temperature increasing from 2K to 6.4 K by 0.4K steps. (b) Scaling analysis of magnetoresistance data. (c) Temperature dependence of the scaling parameter $t$. $T_0$ = 2K is the lowest temperature for the analysis of QCP. }
	\label{2}
\end{figure}

QCP regime ($p = 0.0525$): A long-range antiferromagnetic order eventually stabilizes at the boundary between the superconducting and antiferromagnetic phases at a finite magnetic field. The magnetic field drives a quantum phase transition from long-range superconducting order into long-range antiferromagnetic order, a single QCP dominates this global phase transition, as shown in the orange region of Fig.\ref{5}. This result is consistent with previous reports that QCP behavior was observed at extremely low doping levels in monolayer BSCCO\cite{monolayerBSCCO}. As for the piecewise linear dynamic critical exponent in Fig.\ref{2}(c), it arises from the competition between thermal fluctuation and disorder\cite{DQCPingr,DQCPinOxide}. Thermal fluctuation dominates at relatively high temperatures, leading to $z{\nu}>1$. Quenched disorder has more impact at low temperatures, resulting in $z{\nu}<1$.

\begin{figure}[tb]
	\centering
	\includegraphics[width=1\textwidth]{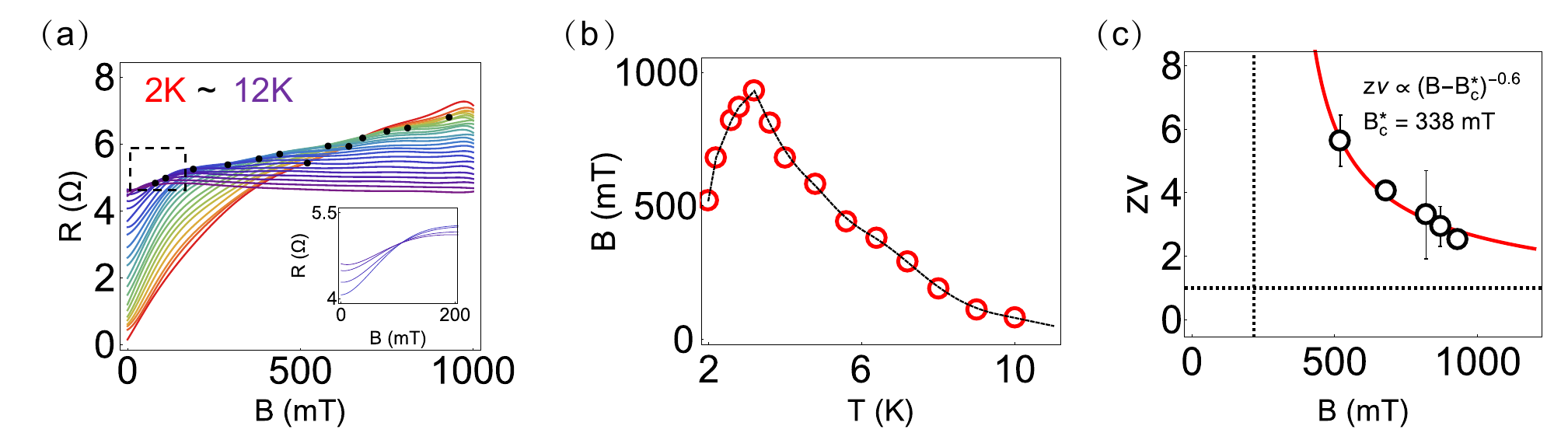}
	\caption{AQGS in BSCCO at p = 0.055. (a) Magneto-resistance with temperature increasing from 2 K to 8 K by 0.4 K steps, and from 8 K to 12 K by 0.5 K steps. The black circles mark the cross points of every two curves at adjacent temperature. The inset shows a zoom-in of data in the vicinity of a typical cross point ( from 9 K to 10.5 K by 0.5 steps), which is marked on the main plot with a dashed box. Details on finite-size scaling analysis for AQGS can be found in the Section 2 of the Supplement Information. (b) Temperature dependence of cross points $B_{c}$. The black curve guides the trend of $B_{c}$ towards the zero temperature.  (c) Dynamical exponent $z{\nu}$ as a function of magnetic field. The red curve shows the fitting from activated scaling law $z{\nu}\propto(B-B_{c}^{*})^{-0.6}$. The horizontal dashed line is $z{\nu} = 1$ and the vertical one marks the extracted $B_{c}^{*}$.}
	\label{3}
\end{figure}

Apart from the study of quantum criticality itself, we can also use quantum criticality behavior to map out a phase diagram of underdoped BSCCO as a function of magnetic field and doping level as summarized in Fig.\ref{5}. Quantum phase transition occurs at zero temperature\cite{ReviewQPT1,ReviewQPT2}. Therefore, to properly map the phase diagram, we have to extract the characteristic magnetic field $B_{c}^{*}$ of quantum criticality at zero temperature from $B_{c}$ measured at finite temperature. This is fairly straightforward in the case of QCP since the critical point is temperature independent. In the case of QGS and AQGS, $B_{c}^{*}$ enclosed in the FSS scaling analysis of critical exponent $z{\nu}$, see vertical dash line in Fig.\ref{3}(c) and Fig.\ref{4}(c), naturally corresponds to the infinite-randomness quantum critical point\cite{QGSinGa,QGSinionicgating,QGSinPdTe,QGSinNiO,AQGS}, and hence labels the  phase boundary. It is interesting to mark that $B_{c}^{*}$ extracted for QGS (circle markers in in Fig.\ref{5}) are in good agreement vortex-melting field at 0 K obtained from RH oscillation measurements (diamond markers)\cite{ourwork}, highlighting the central role of vortex glass state in the case of QGS. From Fig.\ref{5}, it is clear that $B_{c}^{*}$ reduces relatively slowly in QGS regime where the antiferromagnetic correlation is negligible but drops rapidly once the antiferromagnetic correlation is activated in AQGS regime and eventually approaches zero with the establishment of long-range antiferromagnetic order in QCP regime. 

The phase diagram extends the classical view that a single QCP dominates the interaction between superconductivity and other correlated states\cite{QCPinLSCO,monolayerBSCCO,QCPinYBCO} by identifying other quantum criticality behaviors arising from micro-structures\cite{DQCPincuprates,DQCPingr,DQCPinOxide,AQGS}. Especially, it unveils the vital role of AFM micro-structure, including local antiferromagnetic fluctuation or short-range AFM order, in linking SC and AFM and specifies the doping range for the SC-AFM coexisting regime should be $0.0525 < p < 0.06$ in which QCP and AQGS occur. The upper bound of coexisting regime may be underestimated since we cannot exclude the possibility that QGS in the vicinity of $p = 0.06$ may actually be AQGS due to the lack of extremely low temperature data. The additional insights on AFM micro-structure, along with recent works on spin texture\cite{spintexture}, charge-ordered insulator\cite{stmofcuprates} and chessboard-like island\cite{chessboardincuprates}, pinpoints that micro-structures have profound influence in the formation of global superconductivity in high-$T_c$ superconductor.       

In conclusion, we observe an evolution of quantum criticality from QCP to AQGS and eventually ends as QGS with increasing doping levels in underdoped BSCCO. The evolution can be qualitatively captured by the superconducting puddle model with the emergence of antiferromagnetic correlation taken into account. With the evolution of quantum criticality, we can map out a magnetic field-doping level phase diagram for underdoped BSCCO at 0 K. The phase diagram highlights the doping range where antiferromagnetism and superconductivity compete with each other should be $0.0525 < p < 0.06$, and suggests antiferromagnetism manifests itself via local antiferromagnetic fluctuation within normal regions that separate rare superconducting puddles in this regime. Therefore, our results shed light on both quantum criticality in correlated systems and the superconducting pairing mechanism in cuprates. 
\clearpage

\clearpage
\noindent \textbf{Methods}

\noindent\textbf{Single crystals growth.} The growth method of high-quality BSCCO single crystals is the floating zone optical image furnace. Powders of ${\mathrm{Bi}}_{2}{\mathrm{O}}_{3}$, ${\mathrm{Sr}}{\mathrm{CO}}_{3}$, ${\mathrm{Ca}}{\mathrm{CO}}_{3}$, CuO (nominal chemical ratio of Bi:Sr:Ca:Cu=2:2:1:2) were well mixed, grounded and heated at 820 °C for 50 hours  with intermediate grinding. Under the pressure of 70 MPa, the heated powders were pressed into the feed rod. Then, the rod was annealed at 850 $^{\circ}$C for 24 hours in flowing oxygen. The feed rod was melted under a mixed gas flow [${\mathrm{O}}_{2}$ (20$\%$)+Ar (80$\%$)] in the floating zone furnace with a feeding speed of about 0.8 mm/h.

\vspace*{10pt}
\noindent\textbf{Device fabrication and measurements.} BSCCO flakes (30-50 nm) are mechanically exfoliated and dry transferred to pre-patterned electrodes (Ti 15nm/Au 35nm) in a ${\mathrm{N}}_{2}$ filled glove box (${\mathrm{H}}_{2}{\mathrm{O}}<$0.01 ppm, ${\mathrm{O}}_{2}<$0.01 ppm).  The samples are loaded into a high-vacuum cryostat immediately, usually within 30 minutes after exfoliation, to prevent the unintentionally oxygen-releasing process. To carry out 4-probe measurement at different doping levels, we repeat a vacuum annealing measurement protocol: First, the BSCCO sample stays at or above 300 K under a high vacuum condition for a given amount of time. The oxygen is naturally released from the sample, which modulates the doping level. This process is referred to as vacuum annealing. Then, the system is cooled to the base temperature, and detailed measurements are conducted in a 4-probe configuration (with an excitation current of 10 $\mu A$). The oxygen-releasing rate is significantly reduced when the temperature drops below 200 K. We have ensured that the doping level almost remains unchanged in the temperature range of interest ($<$ 100 K) by monitoring the normal state resistance frequently. Relation between the doping level and vacuum annealing time of a typical sample is summarized in Fig.\ref{1}. It is helpful to stress that the relation between doping level and annealing time exhibit sample to sample variation. \\

\clearpage
\bibliography{scibib.bib}

\begin{thebibliography}{10}
\urlstyle{rm}
\expandafter\ifx\csname url\endcsname\relax
  \def\url#1{\texttt{#1}}\fi
\expandafter\ifx\csname urlprefix\endcsname\relax\def\urlprefix{URL }\fi
\expandafter\ifx\csname doiprefix\endcsname\relax\def\doiprefix{DOI: }\fi
\providecommand{\bibinfo}[2]{#2}
\providecommand{\eprint}[2][]{\url{#2}}

\bibitem{IntroFe}
\bibinfo{author}{Fernandes, R.~M.} \emph{et~al.}
\newblock \bibinfo{journal}{\bibinfo{title}{Iron pnictides and chalcogenides: a
  new paradigm for superconductivity}}.
\newblock {\emph{\JournalTitle{Nature}}} \textbf{\bibinfo{volume}{601}},
  \bibinfo{pages}{35--44}, \doiprefix\url{10.1038/s41586-021-04073-2}
  (\bibinfo{year}{2022}).

\bibitem{monolayerBSCCO}
\bibinfo{author}{Yu, Y.} \emph{et~al.}
\newblock \bibinfo{journal}{\bibinfo{title}{High-temperature superconductivity
  in monolayer
  {${\mathrm{Bi}}_{2}{\mathrm{Sr}}_{2}{\mathrm{CaCu}}_{2}{\mathrm{O}}_{8+\ensuremath{\delta}}$}}}.
\newblock {\emph{\JournalTitle{Nature}}} \textbf{\bibinfo{volume}{575}},
  \bibinfo{pages}{156--163}, \doiprefix\url{10.1038/s41586-019-1718-x}
  (\bibinfo{year}{2019}).

\bibitem{QCPinCu2}
\bibinfo{author}{Michon, B.} \emph{et~al.}
\newblock \bibinfo{journal}{\bibinfo{title}{Thermodynamic signatures of quantum
  criticality in cuprate superconductors}}.
\newblock {\emph{\JournalTitle{Nature}}} \textbf{\bibinfo{volume}{567}},
  \bibinfo{pages}{218--222}, \doiprefix\url{10.1038/s41586-019-0932-x}
  (\bibinfo{year}{2019}).

\bibitem{QCPinWTe}
\bibinfo{author}{Song, T.} \emph{et~al.}
\newblock \bibinfo{journal}{\bibinfo{title}{Unconventional superconducting
  quantum criticality in monolayer {${\mathrm{WTe}}_{2}$}}}.
\newblock {\emph{\JournalTitle{Nature Physics}}} \textbf{\bibinfo{volume}{20}},
  \bibinfo{pages}{269--274}, \doiprefix\url{10.1038/s41567-023-02291-1}
  (\bibinfo{year}{2024}).

\bibitem{IntroHF1}
\bibinfo{author}{Gegenwart, P.}, \bibinfo{author}{Si, Q.} \&
  \bibinfo{author}{Steglich, F.}
\newblock \bibinfo{journal}{\bibinfo{title}{Quantum criticality in
  heavy-fermion metals}}.
\newblock {\emph{\JournalTitle{Nature Physics}}} \textbf{\bibinfo{volume}{4}},
  \bibinfo{pages}{186--197}, \doiprefix\url{10.1038/nphys892}
  (\bibinfo{year}{2008}).

\bibitem{IntroHF2}
\bibinfo{author}{Wetli, C.} \emph{et~al.}
\newblock \bibinfo{journal}{\bibinfo{title}{Time-resolved collapse and revival
  of the {Kondo} state near a quantum phase transition}}.
\newblock {\emph{\JournalTitle{Nature Physics}}} \textbf{\bibinfo{volume}{14}},
  \bibinfo{pages}{1103--1107}, \doiprefix\url{10.1038/s41567-018-0228-3}
  (\bibinfo{year}{2018}).

\bibitem{IntroFM1}
\bibinfo{author}{Rowley, S.~E.} \emph{et~al.}
\newblock \bibinfo{journal}{\bibinfo{title}{Ferroelectric quantum
  criticality}}.
\newblock {\emph{\JournalTitle{Nature Physics}}} \textbf{\bibinfo{volume}{10}},
  \bibinfo{pages}{367--372}, \doiprefix\url{10.1038/nphys2924}
  (\bibinfo{year}{2014}).

\bibitem{IntroinTP}
\bibinfo{author}{Ghiotto, A.} \emph{et~al.}
\newblock \bibinfo{journal}{\bibinfo{title}{Quantum criticality in twisted
  transition metal dichalcogenides}}.
\newblock {\emph{\JournalTitle{Nature}}} \textbf{\bibinfo{volume}{597}},
  \bibinfo{pages}{345--349}, \doiprefix\url{10.1038/s41586-021-03815-6}
  (\bibinfo{year}{2021}).

\bibitem{IntroinTP2}
\bibinfo{author}{Li, Q.} \emph{et~al.}
\newblock \bibinfo{journal}{\bibinfo{title}{Tunable quantum criticalities in an
  isospin extended {Hubbard} model simulator}}.
\newblock {\emph{\JournalTitle{Nature}}} \textbf{\bibinfo{volume}{609}},
  \bibinfo{pages}{479--484}, \doiprefix\url{10.1038/s41586-022-05106-0}
  (\bibinfo{year}{2022}).

\bibitem{Intronuclei}
\bibinfo{author}{Cejnar, P.}, \bibinfo{author}{Jolie, J.} \&
  \bibinfo{author}{Casten, R.~F.}
\newblock \bibinfo{journal}{\bibinfo{title}{Quantum phase transitions in the
  shapes of atomic nuclei}}.
\newblock {\emph{\JournalTitle{Rev. Mod. Phys.}}}
  \textbf{\bibinfo{volume}{82}}, \bibinfo{pages}{2155--2212},
  \doiprefix\url{10.1103/RevModPhys.82.2155} (\bibinfo{year}{2010}).

\bibitem{Introdarkmatter}
\bibinfo{author}{Alford, M.~G.}, \bibinfo{author}{Schmitt, A.},
  \bibinfo{author}{Rajagopal, K.} \& \bibinfo{author}{Sch\"afer, T.}
\newblock \bibinfo{journal}{\bibinfo{title}{Color superconductivity in dense
  quark matter}}.
\newblock {\emph{\JournalTitle{Rev. Mod. Phys.}}}
  \textbf{\bibinfo{volume}{80}}, \bibinfo{pages}{1455--1515},
  \doiprefix\url{10.1103/RevModPhys.80.1455} (\bibinfo{year}{2008}).

\bibitem{Strangerthanmetal}
\bibinfo{author}{Phillips, P.~W.}, \bibinfo{author}{Hussey, N.~E.} \&
  \bibinfo{author}{Abbamonte, P.}
\newblock \bibinfo{journal}{\bibinfo{title}{Stranger than metals}}.
\newblock {\emph{\JournalTitle{Science}}} \textbf{\bibinfo{volume}{377}},
  \bibinfo{pages}{{eabh4273}}, \doiprefix\url{10.1126/science.abh4273}
  (\bibinfo{year}{2022}).

\bibitem{Harris_1974}
\bibinfo{author}{Harris, A.~B.}
\newblock \bibinfo{journal}{\bibinfo{title}{Effect of random defects on the
  critical behaviour of {Ising} models}}.
\newblock {\emph{\JournalTitle{Journal of Physics C: Solid State Physics}}}
  \textbf{\bibinfo{volume}{7}}, \bibinfo{pages}{1671},
  \doiprefix\url{10.1088/0022-3719/7/9/009} (\bibinfo{year}{1974}).

\bibitem{harris}
\bibinfo{author}{Vojta, T.} \& \bibinfo{author}{Hoyos, J.~A.}
\newblock \bibinfo{journal}{\bibinfo{title}{Criticality and quenched disorder:
  Harris criterion versus rare regions}}.
\newblock {\emph{\JournalTitle{Phys. Rev. Lett.}}}
  \textbf{\bibinfo{volume}{112}}, \bibinfo{pages}{075702},
  \doiprefix\url{10.1103/PhysRevLett.112.075702} (\bibinfo{year}{2014}).

\bibitem{QCPinInO}
\bibinfo{author}{Hebard, A.~F.} \& \bibinfo{author}{Paalanen, M.~A.}
\newblock \bibinfo{journal}{\bibinfo{title}{Magnetic-field-tuned
  superconductor-insulator transition in two-dimensional films}}.
\newblock {\emph{\JournalTitle{Phys. Rev. Lett.}}}
  \textbf{\bibinfo{volume}{65}}, \bibinfo{pages}{927--930},
  \doiprefix\url{10.1103/PhysRevLett.65.927} (\bibinfo{year}{1990}).

\bibitem{QCPinBi}
\bibinfo{author}{Liu, Y.} \emph{et~al.}
\newblock \bibinfo{journal}{\bibinfo{title}{Scaling of the
  insulator-to-superconductor transition in ultrathin amorphous {Bi} films}}.
\newblock {\emph{\JournalTitle{Phys. Rev. Lett.}}}
  \textbf{\bibinfo{volume}{67}}, \bibinfo{pages}{2068--2071},
  \doiprefix\url{10.1103/PhysRevLett.67.2068} (\bibinfo{year}{1991}).

\bibitem{QCPinLSCO}
\bibinfo{author}{Bollinger, A.~T.} \emph{et~al.}
\newblock \bibinfo{journal}{\bibinfo{title}{Superconductor–insulator
  transition in
  {${\mathrm{La}}_{2-x}{\mathrm{Sr}}_{x}{\mathrm{Cu}}_{2}{\mathrm{O}}_{4}$} at
  the pair quantum resistance}}.
\newblock {\emph{\JournalTitle{Nature}}} \textbf{\bibinfo{volume}{472}},
  \bibinfo{pages}{458--460}, \doiprefix\url{10.1038/nature09998}
  (\bibinfo{year}{2011}).

\bibitem{QGSinGa}
\bibinfo{author}{Xing, Y.} \emph{et~al.}
\newblock \bibinfo{journal}{\bibinfo{title}{Quantum {Griffiths} singularity of
  superconductor-metal transition in {Ga} thin films}}.
\newblock {\emph{\JournalTitle{Science}}} \textbf{\bibinfo{volume}{350}},
  \bibinfo{pages}{542--545}, \doiprefix\url{10.1126/science.aaa7154}
  (\bibinfo{year}{2015}).

\bibitem{QGSinionicgating}
\bibinfo{author}{Saito, Y.}, \bibinfo{author}{Nojima, T.} \&
  \bibinfo{author}{Iwasa, Y.}
\newblock \bibinfo{journal}{\bibinfo{title}{Quantum phase transitions in highly
  crystalline two-dimensional superconductors}}.
\newblock {\emph{\JournalTitle{Nature Communications}}}
  \textbf{\bibinfo{volume}{9}}, \bibinfo{pages}{778},
  \doiprefix\url{10.1038/s41467-018-03275-z} (\bibinfo{year}{2018}).

\bibitem{QGSinPdTe}
\bibinfo{author}{Liu, Y.} \emph{et~al.}
\newblock \bibinfo{journal}{\bibinfo{title}{Observation of in-plane quantum
  {Griffiths} singularity in two-dimensional crystalline superconductors}}.
\newblock {\emph{\JournalTitle{Phys. Rev. Lett.}}}
  \textbf{\bibinfo{volume}{127}}, \bibinfo{pages}{137001},
  \doiprefix\url{10.1103/PhysRevLett.127.137001} (\bibinfo{year}{2021}).

\bibitem{QGSinNiO}
\bibinfo{author}{Zhao, Q.} \emph{et~al.}
\newblock \bibinfo{journal}{\bibinfo{title}{Isotropic quantum {Griffiths}
  singularity in {${\mathrm{Nd}}_{0.8}{\mathrm{Sr}}_{0.2}{\mathrm{NiO}}_{2}$}
  infinite-layer superconducting thin films}}.
\newblock {\emph{\JournalTitle{Phys. Rev. Lett.}}}
  \textbf{\bibinfo{volume}{133}}, \bibinfo{pages}{036003},
  \doiprefix\url{10.1103/PhysRevLett.133.036003} (\bibinfo{year}{2024}).

\bibitem{AQGS}
\bibinfo{author}{Liu, Y.} \emph{et~al.}
\newblock \bibinfo{journal}{\bibinfo{title}{Anomalous quantum {Griffiths}
  singularity in ultrathin crystalline lead films}}.
\newblock {\emph{\JournalTitle{Nature Communications}}}
  \textbf{\bibinfo{volume}{10}}, \bibinfo{pages}{3633},
  \doiprefix\url{10.1038/s41467-019-11607-w} (\bibinfo{year}{2019}).

\bibitem{AQGS2}
\bibinfo{author}{Wang, B.} \emph{et~al.}
\newblock \bibinfo{journal}{\bibinfo{title}{Effectively tuning the quantum
  griffiths phase by controllable quantum fluctuations}}.
\newblock {\emph{\JournalTitle{Science Advances}}}
  \textbf{\bibinfo{volume}{10}}, \bibinfo{pages}{eadp1402},
  \doiprefix\url{10.1126/sciadv.adp1402} (\bibinfo{year}{2024}).

\bibitem{IonicBSCCO}
\bibinfo{author}{Wang, T.} \emph{et~al.}
\newblock \bibinfo{journal}{\bibinfo{title}{Universal relation between doping
  content and normal-state resistance in gate voltage tuned ultrathin
  {${\mathrm{Bi}}_{2}{\mathrm{Sr}}_{2}{\mathrm{CaCu}}_{2}{\mathrm{O}}_{8+x}$}
  flakes}}.
\newblock {\emph{\JournalTitle{Phys. Rev. B}}} \textbf{\bibinfo{volume}{106}},
  \bibinfo{pages}{104509}, \doiprefix\url{10.1103/PhysRevB.106.104509}
  (\bibinfo{year}{2022}).

\bibitem{zvinBi}
\bibinfo{author}{Markovi$\'{c}$, N.}, \bibinfo{author}{Christiansen, C.} \&
  \bibinfo{author}{Goldman, A.~M.}
\newblock \bibinfo{journal}{\bibinfo{title}{Thickness--magnetic field phase
  diagram at the superconductor-insulator transition in {2D}}}.
\newblock {\emph{\JournalTitle{Phys. Rev. Lett.}}}
  \textbf{\bibinfo{volume}{81}}, \bibinfo{pages}{5217--5220},
  \doiprefix\url{10.1103/PhysRevLett.81.5217} (\bibinfo{year}{1998}).

\bibitem{zvinPCCO}
\bibinfo{author}{Zeng, S.~W.} \emph{et~al.}
\newblock \bibinfo{journal}{\bibinfo{title}{Two-dimensional
  superconductor-insulator quantum phase transitions in an electron-doped
  cuprate}}.
\newblock {\emph{\JournalTitle{Phys. Rev. B}}} \textbf{\bibinfo{volume}{92}},
  \bibinfo{pages}{020503}, \doiprefix\url{10.1103/PhysRevB.92.020503}
  (\bibinfo{year}{2015}).

\bibitem{goldman}
\bibinfo{author}{Lewellyn, N.~A.} \emph{et~al.}
\newblock \bibinfo{journal}{\bibinfo{title}{Infinite-randomness fixed point of
  the quantum superconductor-metal transitions in amorphous thin films}}.
\newblock {\emph{\JournalTitle{Phys. Rev. B}}} \textbf{\bibinfo{volume}{99}},
  \bibinfo{pages}{054515}, \doiprefix\url{10.1103/PhysRevB.99.054515}
  (\bibinfo{year}{2019}).

\bibitem{DQCPingr}
\bibinfo{author}{Sun, Y.} \emph{et~al.}
\newblock \bibinfo{journal}{\bibinfo{title}{Double quantum criticality in
  superconducting tin arrays-graphene hybrid}}.
\newblock {\emph{\JournalTitle{Nature Communications}}}
  \textbf{\bibinfo{volume}{9}}, \bibinfo{pages}{2159},
  \doiprefix\url{10.1038/s41467-018-04606-w} (\bibinfo{year}{2018}).

\bibitem{QGSingr}
\bibinfo{author}{Chen, F.} \emph{et~al.}
\newblock \bibinfo{journal}{\bibinfo{title}{Quantum {Griffiths} singularity in
  ordered artificial superconducting-islands-array on graphene}}.
\newblock {\emph{\JournalTitle{Nano Letters}}} \textbf{\bibinfo{volume}{24}},
  \bibinfo{pages}{2444--2450}, \doiprefix\url{10.1021/acs.nanolett.3c03870}
  (\bibinfo{year}{2024}).

\bibitem{reviewofQGS}
\bibinfo{author}{Wang, Z.}, \bibinfo{author}{Liu, Y.}, \bibinfo{author}{Ji, C.}
  \& \bibinfo{author}{Wang, J.}
\newblock \bibinfo{journal}{\bibinfo{title}{Quantum phase transitions in
  two-dimensional superconductors: a review on recent experimental progress}}.
\newblock {\emph{\JournalTitle{Reports on Progress in Physics}}}
  \textbf{\bibinfo{volume}{87}}, \bibinfo{pages}{014502},
  \doiprefix\url{10.1088/1361-6633/ad14f3} (\bibinfo{year}{2023}).

\bibitem{ReviewQPT2}
\bibinfo{author}{Vojta, M.}
\newblock \bibinfo{journal}{\bibinfo{title}{Quantum phase transitions}}.
\newblock {\emph{\JournalTitle{Reports on Progress in Physics}}}
  \textbf{\bibinfo{volume}{66}}, \bibinfo{pages}{2069},
  \doiprefix\url{10.1088/0034-4885/66/12/R01} (\bibinfo{year}{2003}).

\bibitem{Lin1}
\bibinfo{author}{Lin, Y.-H.}, \bibinfo{author}{Nelson, J.} \&
  \bibinfo{author}{Goldman, A.~M.}
\newblock \bibinfo{journal}{\bibinfo{title}{Suppression of the
  berezinskii-kosterlitz-thouless transition in 2d superconductors by
  macroscopic quantum tunneling}}.
\newblock {\emph{\JournalTitle{Phys. Rev. Lett.}}}
  \textbf{\bibinfo{volume}{109}}, \bibinfo{pages}{017002},
  \doiprefix\url{10.1103/PhysRevLett.109.017002} (\bibinfo{year}{2012}).

\bibitem{Lin2}
\bibinfo{author}{Lin, Y.-H.} \& \bibinfo{author}{Goldman, A.~M.}
\newblock \bibinfo{journal}{\bibinfo{title}{Magnetic-field-tuned quantum phase
  transition in the insulating regime of ultrathin amorphous bi films}}.
\newblock {\emph{\JournalTitle{Phys. Rev. Lett.}}}
  \textbf{\bibinfo{volume}{106}}, \bibinfo{pages}{127003},
  \doiprefix\url{10.1103/PhysRevLett.106.127003} (\bibinfo{year}{2011}).

\bibitem{ourwork}
\bibinfo{author}{Lan, C.} \emph{et~al.}
\newblock \bibinfo{journal}{\bibinfo{title}{Vortex-driven periodic and
  aperiodic magnetoresistance oscillations in cuprates}}.
\newblock {\emph{\JournalTitle{Phys. Rev. B}}} \textbf{\bibinfo{volume}{109}},
  \bibinfo{pages}{115133}, \doiprefix\url{10.1103/PhysRevB.109.115133}
  (\bibinfo{year}{2024}).

\bibitem{reentrant}
\bibinfo{author}{Zhao, Y.} \emph{et~al.}
\newblock \bibinfo{journal}{\bibinfo{title}{Normal-state reentrant behavior in
  superconducting
  {${\mathrm{Bi}}_{2}$${\mathrm{Sr}}_{2}$${\mathrm{CaCu}}_{2}$${\mathrm{O}}_{8}$$\setminus$
  ${\mathrm{Bi}}_{2}$${\mathrm{Sr}}_{2}$${\mathrm{Ca}}_{2}$${\mathrm{Cu}}_{3}$${\mathrm{O}}_{10}$}
  intergrowth single crystals}}.
\newblock {\emph{\JournalTitle{Phys. Rev. B}}} \textbf{\bibinfo{volume}{51}},
  \bibinfo{pages}{3134--3139}, \doiprefix\url{10.1103/PhysRevB.51.3134}
  (\bibinfo{year}{1995}).

\bibitem{SITofBSCCO}
\bibinfo{author}{Liao, M.} \emph{et~al.}
\newblock \bibinfo{journal}{\bibinfo{title}{Superconductor–insulator
  transitions in exfoliated
  {${\mathrm{Bi}}_{2}{\mathrm{Sr}}_{2}{\mathrm{CaCu}}_{2}{\mathrm{O}}_{8+\ensuremath{\delta}}$}
  flakes}}.
\newblock {\emph{\JournalTitle{Nano Letters}}} \textbf{\bibinfo{volume}{18}},
  \bibinfo{pages}{5660--5665}, \doiprefix\url{10.1021/acs.nanolett.8b02183}
  (\bibinfo{year}{2018}).

\bibitem{DQCPinOxide}
\bibinfo{author}{Biscaras, J.} \emph{et~al.}
\newblock \bibinfo{journal}{\bibinfo{title}{Multiple quantum criticality in a
  two-dimensional superconductor}}.
\newblock {\emph{\JournalTitle{Nature Materials}}}
  \textbf{\bibinfo{volume}{12}}, \bibinfo{pages}{542--548},
  \doiprefix\url{10.1038/nmat3624} (\bibinfo{year}{2013}).

\bibitem{SOCincuprates}
\bibinfo{author}{Jiang, K.}, \bibinfo{author}{Wu, X.}, \bibinfo{author}{Hu, J.}
  \& \bibinfo{author}{Wang, Z.}
\newblock \bibinfo{journal}{\bibinfo{title}{Nodeless high-{${\mathrm{T}}_{c}$}
  superconductivity in the highly overdoped {${\mathrm{CuO}}_{2}$} monolayer}}.
\newblock {\emph{\JournalTitle{Phys. Rev. Lett.}}}
  \textbf{\bibinfo{volume}{121}}, \bibinfo{pages}{227002},
  \doiprefix\url{10.1103/PhysRevLett.121.227002} (\bibinfo{year}{2018}).

\bibitem{afmwithsc}
\bibinfo{author}{Eisaki, H.} \emph{et~al.}
\newblock \bibinfo{journal}{\bibinfo{title}{Competition between magnetism and
  superconductivity in rare-earth nickel boride carbides}}.
\newblock {\emph{\JournalTitle{Phys. Rev. B}}} \textbf{\bibinfo{volume}{50}},
  \bibinfo{pages}{647--650}, \doiprefix\url{10.1103/PhysRevB.50.647}
  (\bibinfo{year}{1994}).

\bibitem{puddle2}
\bibinfo{author}{Fratini, M.} \emph{et~al.}
\newblock \bibinfo{journal}{\bibinfo{title}{Scale-free structural organization
  of oxygen interstitials in la2cuo4+y}}.
\newblock {\emph{\JournalTitle{Nature}}} \textbf{\bibinfo{volume}{466}},
  \bibinfo{pages}{841--844}, \doiprefix\url{10.1038/nature09260}
  (\bibinfo{year}{2010}).

\bibitem{monolayerBSCCO2}
\bibinfo{author}{Wang, S.} \emph{et~al.}
\newblock \bibinfo{journal}{\bibinfo{title}{Oscillating paramagnetic meissner
  effect and berezinskii-kosterlitz-thouless transition in underdoped
  ${\mathrm{bi}}_{2}{\mathrm{sr}}_{2}{\mathrm{cacu}}_{2}{\mathrm{o}}_{8+\ensuremath{\delta}}$}}.
\newblock {\emph{\JournalTitle{National Science Review}}}
  \textbf{\bibinfo{volume}{11}}, \bibinfo{pages}{nwad249},
  \doiprefix\url{10.1093/nsr/nwad249} (\bibinfo{year}{2023}).

\bibitem{stmofcuprates}
\bibinfo{author}{Cai, P.} \emph{et~al.}
\newblock \bibinfo{journal}{\bibinfo{title}{Visualizing the evolution from the
  {Mott} insulator to a charge-ordered insulator in lightly doped cuprates}}.
\newblock {\emph{\JournalTitle{Nature Physics}}} \textbf{\bibinfo{volume}{12}},
  \bibinfo{pages}{1047--1051}, \doiprefix\url{10.1038/nphys3840}
  (\bibinfo{year}{2016}).

\bibitem{puddle1}
\bibinfo{author}{Spivak, B.}, \bibinfo{author}{Oreto, P.} \&
  \bibinfo{author}{Kivelson, S.~A.}
\newblock \bibinfo{journal}{\bibinfo{title}{Theory of quantum metal to
  superconductor transitions in highly conducting systems}}.
\newblock {\emph{\JournalTitle{Phys. Rev. B}}} \textbf{\bibinfo{volume}{77}},
  \bibinfo{pages}{214523}, \doiprefix\url{10.1103/PhysRevB.77.214523}
  (\bibinfo{year}{2008}).

\bibitem{afmincuprates2}
\bibinfo{author}{Kurokawa, K.} \emph{et~al.}
\newblock \bibinfo{journal}{\bibinfo{title}{Unveiling phase diagram of the
  lightly doped high-{$T_{c}$} cuprate superconductors with disorder removed}}.
\newblock {\emph{\JournalTitle{Nature Communications}}}
  \textbf{\bibinfo{volume}{14}}, \bibinfo{pages}{4064},
  \doiprefix\url{10.1038/s41467-023-39457-7} (\bibinfo{year}{2023}).

\bibitem{afmincuprates}
\bibinfo{author}{Lavrov, A.~N.}, \bibinfo{author}{Kozeeva, L.~P.},
  \bibinfo{author}{Trunin, M.~R.} \& \bibinfo{author}{Zverev, V.~N.}
\newblock \bibinfo{journal}{\bibinfo{title}{Competition and coexistence of
  antiferromagnetism and superconductivity in
  {$R{\text{Ba}}_{2}{\text{Cu}}_{3}{\text{O}}_{6+x}\text{ }(R=\text{Lu},\text{
  }\text{Y})$} single crystals}}.
\newblock {\emph{\JournalTitle{Phys. Rev. B}}} \textbf{\bibinfo{volume}{79}},
  \bibinfo{pages}{214523}, \doiprefix\url{10.1103/PhysRevB.79.214523}
  (\bibinfo{year}{2009}).

\bibitem{localafm3}
\bibinfo{author}{Chang, J.} \emph{et~al.}
\newblock \bibinfo{journal}{\bibinfo{title}{Tuning competing orders in
  {${\text{La}}_{2\ensuremath{-}x}{\text{Sr}}_{x}{\text{CuO}}_{4}$} cuprate
  superconductors by the application of an external magnetic field}}.
\newblock {\emph{\JournalTitle{Phys. Rev. B}}} \textbf{\bibinfo{volume}{78}},
  \bibinfo{pages}{104525}, \doiprefix\url{10.1103/PhysRevB.78.104525}
  (\bibinfo{year}{2008}).

\bibitem{ReviewQPT1}
\bibinfo{author}{Sondhi, S.~L.}, \bibinfo{author}{Girvin, S.~M.},
  \bibinfo{author}{Carini, J.~P.} \& \bibinfo{author}{Shahar, D.}
\newblock \bibinfo{journal}{\bibinfo{title}{Continuous quantum phase
  transitions}}.
\newblock {\emph{\JournalTitle{Rev. Mod. Phys.}}}
  \textbf{\bibinfo{volume}{69}}, \bibinfo{pages}{315--333},
  \doiprefix\url{10.1103/RevModPhys.69.315} (\bibinfo{year}{1997}).

\bibitem{QCPinYBCO}
\bibinfo{author}{Seidler, G.~T.}, \bibinfo{author}{Rosenbaum, T.~F.} \&
  \bibinfo{author}{Veal, B.~W.}
\newblock \bibinfo{journal}{\bibinfo{title}{Two-dimensional
  superconductor-insulator transition in bulk single-crystal
  {${\mathrm{YBa}}_{2}$${\mathrm{Cu}}_{3}$${\mathrm{O}}_{6.38}$}}}.
\newblock {\emph{\JournalTitle{Phys. Rev. B}}} \textbf{\bibinfo{volume}{45}},
  \bibinfo{pages}{10162--10164}, \doiprefix\url{10.1103/PhysRevB.45.10162}
  (\bibinfo{year}{1992}).

\bibitem{DQCPincuprates}
\bibinfo{author}{Shi, X.}, \bibinfo{author}{Lin, P.~V.},
  \bibinfo{author}{Sasagawa, T.}, \bibinfo{author}{Dobrosavljević, V.} \&
  \bibinfo{author}{Popović, D.}
\newblock \bibinfo{journal}{\bibinfo{title}{Two-stage magnetic-field-tuned
  superconductor–insulator transition in underdoped
  {${\mathrm{La}}_{2-x}{\mathrm{Sr}}_{x}{\mathrm{Cu}}_{2}{\mathrm{O}}_{4}$}}}.
\newblock {\emph{\JournalTitle{Nature Physics}}} \textbf{\bibinfo{volume}{10}},
  \bibinfo{pages}{437--443}, \doiprefix\url{10.1038/nphys2961}
  (\bibinfo{year}{2014}).

\bibitem{spintexture}
\bibinfo{author}{Wang, Z.} \emph{et~al.}
\newblock \bibinfo{journal}{\bibinfo{title}{Topological spin texture in the
  pseudogap phase of a high-{$T_{c}$} superconductor}}.
\newblock {\emph{\JournalTitle{Nature}}} \textbf{\bibinfo{volume}{615}},
  \bibinfo{pages}{405--410}, \doiprefix\url{10.1038/s41586-023-05731-3}
  (\bibinfo{year}{2023}).

\bibitem{chessboardincuprates}
\bibinfo{author}{Ye, S.} \emph{et~al.}
\newblock \bibinfo{journal}{\bibinfo{title}{The emergence of global phase
  coherence from local pairing in underdoped cuprates}}.
\newblock {\emph{\JournalTitle{Nature Physics}}} \textbf{\bibinfo{volume}{19}},
  \bibinfo{pages}{1301--1307}, \doiprefix\url{10.1038/s41567-023-02100-9}
  (\bibinfo{year}{2023}).

\end{thebibliography}

\vspace*{10pt}
\noindent \textbf{Acknowledgements} 

\noindent We thank X. Liu for fruitful discussions. We acknowledge the support from the National Natural Science Foundation of China (12204184, 12074134), and the Fundamental Research Funds for the Central Universities (YCJJ20242113).\\

\noindent \textbf{Author contributions} 

\noindent  C. L., C. Y. and S. W. conceived and designed the experiment. C. L. fabricated the samples. C. L. and C.Y. conducted experiments with technique supports from Y. L., Q. C., H. G., and X. Z.. C. L. and Y. L.  performed the scaling analysis. D. W. grew the single crystals. B. Z. and Y. Z. performed spectroscopic characterization of the single crystals. C.L., C.Y. and S. W. analyzed the data and wrote the manuscript. C.Y. and S. W. supervised the project. All authors discussed the results and contributed to the manuscript.\\

\noindent \textbf{Competing interests} 

\noindent The authors declare no competing interests. \\

\noindent \textbf{Data availability}

\noindent The data that support the findings of this study are available from the corresponding author upon reasonable request. \\

\noindent \textbf{Code availability}

\noindent The code that support the findings of this study are available from the corresponding author upon reasonable request. \\

\clearpage 

\clearpage

\clearpage

\clearpage

\clearpage

\clearpage

\end{document}